\documentclass[pdflatex,sn-mathphys]{sn-jnl}% Math and Physical Sciences Reference Style
\usepackage{microtype}
\usepackage{subcaption}
\usepackage[version=4]{mhchem}
\usepackage{siunitx}
\usepackage{bookmark}

\DeclareSIUnit\molar{\mole\per\cubic\deci\metre}
\DeclareSIUnit\Molar{mol/L}

\jyear{2021}%

\renewcommand{\paragraph}[1]{}

\begin{document}

\title[autoSDC]{Towards automated design of corrosion resistant alloy coatings with an autonomous scanning droplet cell}

\author*[1]{\fnm{Brian} \sur{DeCost (0000-0002-3459-5888)}}\email{brian.decost@nist.gov}
\equalcont{These authors contributed equally to this work.}

\author[1]{\fnm{Howie} \sur{Joress (0000-0002-6552-2972)}}\email{howie.joress@nist.gov}
\equalcont{These authors contributed equally to this work.}

\author[2]{\fnm{Suchismita} \sur{Sarker (0000-0002-8820-1143)}}\email{sarker@slac.stanford.edu}
\author[2]{\fnm{Apurva} \sur{Mehta (0000-0003-0870-6932)}}\email{mehta@slac.stanford.edu}

\author[3]{\fnm{Jason} \sur{Hattrick-Simpers (0000-0003-2937-3188)}}\email{jason.hattrick.simpers@utoronto.ca}

\affil*[1]{\orgdiv{Material Measurement Lab}, \orgname{NIST}, \orgaddress{ \city{Gaithersburg},  \state{MD}, \country{USA}}}

\affil[2]{\orgname{SLAC National Accelerator Laboratory}, \orgaddress{\city{Menlo Park}, \state{CA}, \country{USA}}}

\affil[3]{\orgdiv{Dept. of Materials Science}, \orgname{University of Toronto}, \orgaddress{\city{Toronto}, \state{ON}, \country{Canada}}}


\abstract{

We present an autonomous scanning droplet cell platform designed for on-demand alloy electrodeposition and real-time electrochemical characterization for investigating the corrosion-resistance properties of multicomponent alloys.
Automation and machine learning are currently driving rapid innovation in high throughput and autonomous materials design and discovery.
We present two alloy design case studies: one focusing on a multi-objective corrosion resistant alloy optimization, and a case study highlighting the complexity of the multimodal characterization needed to provide insight into the underlying structural and chemical factors that drive observed material behavior.
This motivates a close coupling between autonomous research platforms and scientific machine learning methodology that blends mechanistic physical models and black box machine learning models.
This emerging research area presents new opportunities to accelerate materials synthesis, evaluation, and hence discovery and design.
}




\maketitle

\section{Introduction}\label{sec:intro}

\paragraph{Machine learning and automation.}
Integrating experiments and theory-based modeling to understand and design complex multiscale materials is an important~\cite{hamming1986} and perennial research topic in materials science and engineering~\cite{Allison2006,icme-report,holdren2014}.
The recent rekindling of interest in applications of machine learning (ML) to materials science problems~\cite{Dimiduk2018} has intensified this focus on integrating computation and experiments.
This is clearly illustrated in the nascent autonomous materials science and discovery community~\cite{MAP,Montoya2020,stach2021autonomous}, which has a strong focus on orchestrating automated experimental and computational materials science experiments through adaptive machine learning systems for data evaluation and experimental planning.

\paragraph{Themes in literature}
The fields of active learning and design of experiments uses ML models to inform a series of decisions about which experiments may be valuable to perform in the lab~\cite{lookman2016,rohr2020benchmarking} or through on-demand physics based modeling of  alloys~\cite{Talapatra2018}, condensed-phase materials~\cite{dunn2019rocketsled}, and molecules~\cite{SanchezLengeling2018}.
The autonomous materials experimentation community is consistently demonstrating the power of fully closing the synthesize--characterize--predict loop across a diverse and growing portfolio of materials technologies.
Notable examples include optimizing the growth rate, and thereby quality, of carbon nanotubes by tuning chemical vapor deposition conditions~\cite{nikolaev16_auton_mater_resear}, optimizing the strength of three-dimensional geometry of additively manufactured polymer lattices by varying their superstructure~\cite{gongora2020bayesian}, and autonomously mapping out non-equilibrium processing phase diagrams~\cite{Ament2021}.

In addition to directly optimizing material properties, there is promising progress in roboticization for learning to fabricate materials and chemicals with desired structure, and for efficient on-demand acquisition of complex and/or expensive experimental data.
For example, synchrotron X-ray measurements of nanoparticle density \cite{Noack2019}, and accelerated structural phase map acquisition~\cite{Kusne2020}.

\paragraph{Looking forward: jointly learning complex interdependencies between structure, processing, and properties}
This manuscript explores some of the most challenging aspects of closed loop materials design:
Material performance is linked to process and composition via complex relationships, mediated by structure and chemistry across multiple length scales, such as phase distribution, density, microstructure, and surface quality.
Integrating enough materials synthesis and characterization capabilities into experimental systems to address such complex materials phenomena is a substantial challenge.  A major roadblock to incorporating this complexity into autonomous materials research platforms is addressing the need for quantitative data analysis at scale, which with the advent of high throughput material synthesis and measurement systems is often a primary research bottleneck~\cite{Umehara2019}.
This is an opportunity for mutual progress in autonomous materials research and scientific machine learning~\cite{baker2019workshop}, which is the subfield of ML concerned with designing models and algorithms imbued with physics-based bias and model structure.
Early materials research in this area focused on explicit incorporation of domain knowledge into model form~\cite{childs2019embedding,ren2020embedding,pun2019} and  incorporating automated reasoning and physical constraints into machine learning algorithms~\cite{Gomes2019,Chen2021}. 
A particularly important benefit of incorporating explicit physical insight into  ML models is the potential for substantial efficiency improvements by directly targeting scientific goals through the acquisition policy of an active learning system, for example in measuring a magnetic transition temperature via neutron diffraction~\cite{mcdannald2021fly}.

% - performing well with experimental uncertainty \\
% - NLP oriented: identifying recipes from the literature~\cite{Kim2017,kononova2021opportunities,Mehr2020}. \\

\paragraph{making stuff}
Often, even once desirable composition and structural features are known, the path to actually fabricating usable material with these properties is unclear, presenting an opportunity for creative automated systems to have high impact.
For example, the Chemputer project~\cite{Steiner2019,Gromski2020,Angelone2020} aims to fully automate organic synthetic chemistry with general-purpose hardware, and to algorithmically discover efficient synthesis routes any target molecules.
Focused on metals, Ref. \cite{Boyce2019} outlines a diverse set of synthetic approaches spanning composition and thermomechanical gradient techniques, batch casting and additive manufacturing methods~\cite{Deneault2021}, and roboticized microscopy platforms.
For addressing complex materials systems dominated by multiscale structure, modular clusters of multiple synthesis and characterization tools~\cite{dippo2021highthroughput} may provide a path forward.

\paragraph{Context for SDC work}
This manuscript explores these issues through the context of an automated alloy electrodeposition and characterization platform for alloy design.
A wide variety of high throughput experiment approaches for electrochemical systems have been explored~\cite{muster2011review}, including scanning droplet cell (SDC) systems~\cite{gregoire2013scanning, klemm2011combinatorial,klemm2011high}.

% - demonstrate multiobjective corrosion resistance assay, discuss considerations on materials design and development

\section{Methods}\label{sec:methods}
\subsection{Automated Scanning Droplet Cell}

\begin{figure}[h!btp]
    \centering
    \begin{subfigure}[b]{0.45\textwidth}
         \centering
         \includegraphics[width=\textwidth]{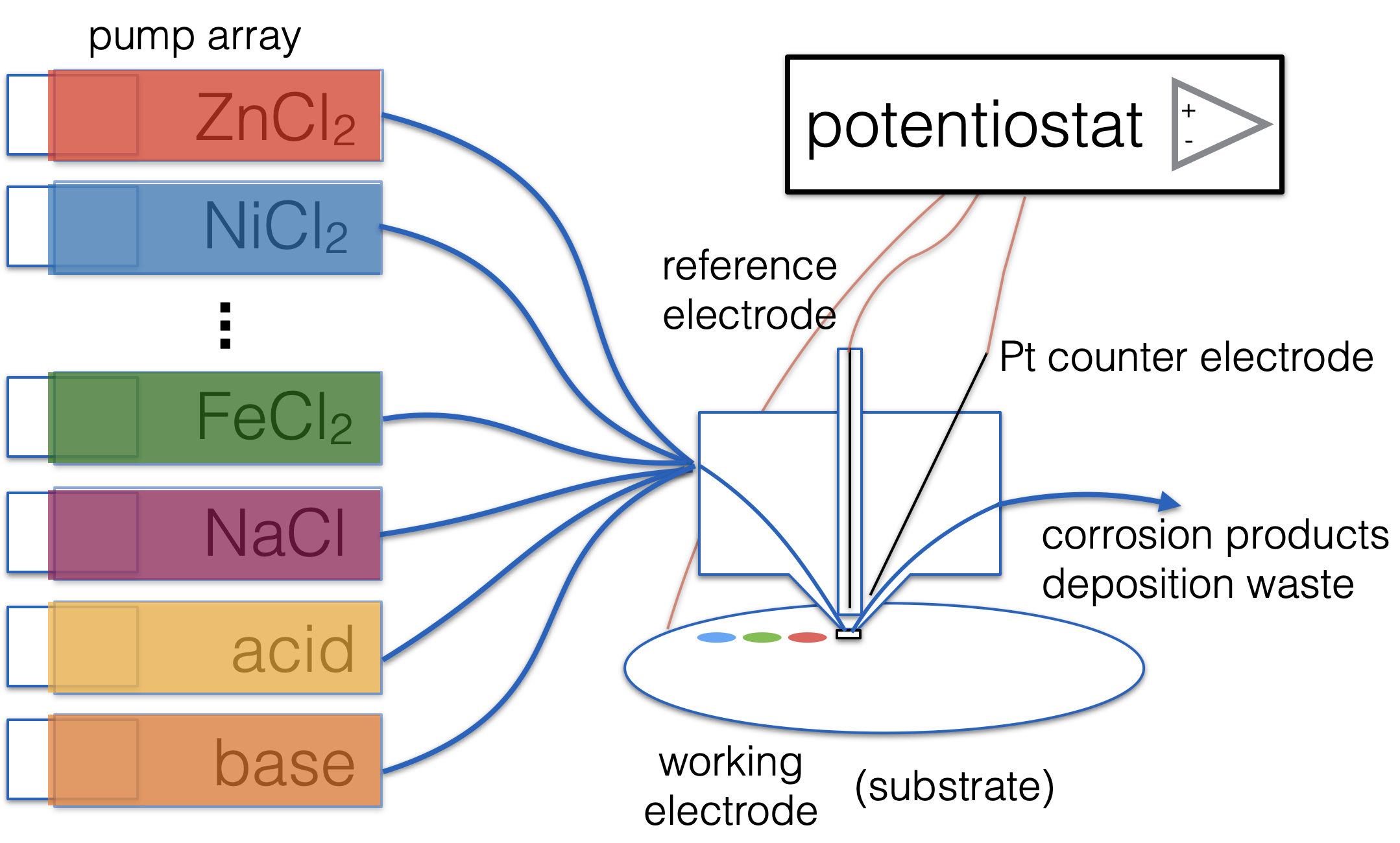}
         \caption{SDC schematic}
         \label{fig:sdc:schematic}
     \end{subfigure}
     \hfill
     \begin{subfigure}[b]{0.45\textwidth}
         \centering
         \includegraphics[width=\textwidth]{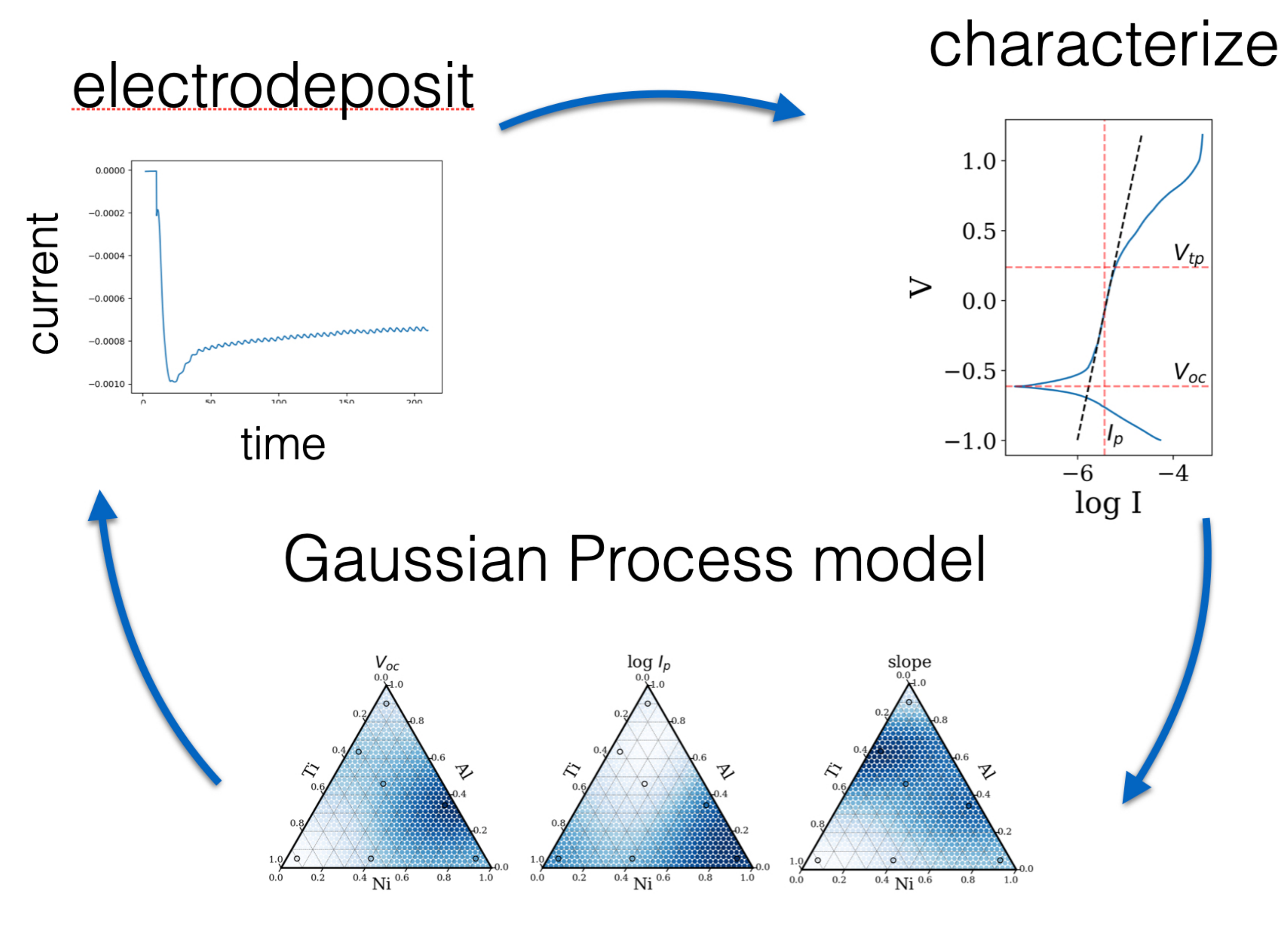}
         \caption{active learning loop}
         \label{fig:loop}
     \end{subfigure}
    \caption{(\subref{fig:sdc:schematic}) Schematic design of the automated scanning droplet cell platform. (\subref{fig:loop}) Schematic joint alloy deposition and property optimization loop.}
    \label{fig:sdc}
\end{figure}

Our high throughput scanning droplet cell platform~\cite{sdcplatform} (illustrated in Figure~\ref{fig:sdc:schematic}) integrates a compact electrochemical flow cell with a fully automated bank of syringe pumps and scanning sample stage.
The flow cell defines a roughly 4.5~\si{mm} diameter circular footprint on the active surface.
The cell can be addressed to any point across a planar sample (\textit{e.g.,} a uniform or composition spread thin film, or a bulk alloy sample) and conduct an agile automated serial electrodeposition and/or corrosion experiments under a variety of solution conditions.
As such, this tool enables online optimization of electrodeposited alloy composition by programatically adjusting solution chemistry and applied potential or current (as briefly discussed in section \ref{sec:discussion}).  In addition, corrosion assays can be performed as a function of composition or structure on these electrodeposited samples or (as in Section \ref{sec:multiopt}) with composition gradient thin film and bulk samples.  Rapid characterization of potential-pH behavior~\cite{sdcplatform} can also be characterized.

Figure~\ref{fig:loop} illustrates the core adaptive alloy electroplating optimization loop.
An initial alloy thin film is deposited with candidate settings for solution composition and deposition conditions (applied current, flow rate, etc.).
Online process monitoring of variables including measured potential, current, pH, and temperature can provide early indication of automation failure or infeasibility of the candidate process settings.

\begin{figure}[h!btp]
    \centering
    \includegraphics[width=0.5\textwidth]{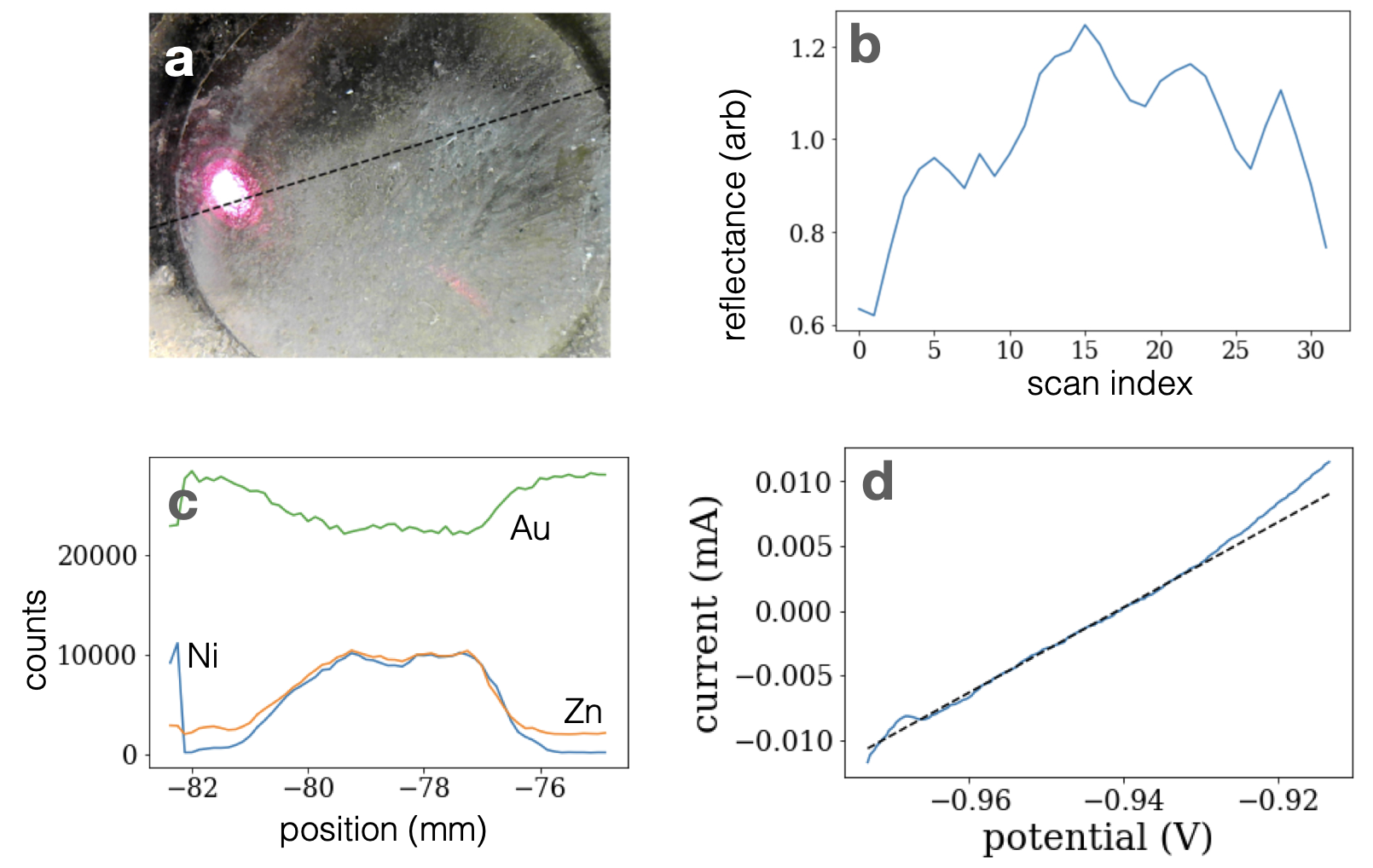}
    \caption{a) surface inspection image b) laser reflectance line scan c) XRF linescan  d) Linear Polarization Resistance}
    \label{fig:BNL:setup}
\end{figure}

Next, a central challenge of building autonomous materials research platforms: integrating as much online characterization capability as possible.
Our system currently performs routine macroscopic surface image acquisition via optical camera  (Figure~\ref{fig:BNL:setup}a) and laser (635 nm wavelength) reflectance line scans (Figure~\ref{fig:BNL:setup}b) to assess the continuity, coloration, uniformity, and qualitative roughness of electrodeposits.
Our platform has a modular design, allowing it to readily be rebuilt around and incorporated into a synchrotron measurement station, enabling online acquisition of \textit{e.g.,} X-ray fluorescence (Figure~\ref{fig:BNL:setup}c), diffraction, and absorption spectroscopy data.

After any online chemical and structural characterization, a wide variety of corrosion resistance assays are possible.
The corrosion environment can be tailored along pH and composition axes through flow mixing, and in principle a dynamically specified series of electrochemical experiments can be performed.

Finally, the experimental loop is closed by linking processing variables and any measured chemical, structural, and performance quantities through a probabilistic machine learning system, such as the Gaussian Process model linking composition to multiple corrosion figures of merit illustrated in Figure~\ref{fig:loop}.
Active learning algorithms use the model predictions and associated uncertainties to prioritize subsequent candidate experiments so as to maximize the probability of learning valuable information towards achieving a design or discovery goal.

% \subsection{Automation setup}

\subsection{Modeling}
\paragraph{Gaussian Process}
Throughout our work, we use the Gaussian Process (GP)~\cite{rw} modeling framework for its flexibility, intrinsic treatment of predictive uncertainty, and facility for automatic hyperparameter optimization.
GPs are Bayesian machine learning models that are similar in spirit to a kernel version of Bayesian linear regression; for a comprehensive and accessible introduction, refer to~\citep{Grtler2019}.
GP model specification is often described in terms of building a Bayesian prior over functions that might possibly describe the data.
This is decomposed into two elements: the mean of the function prior, $m(\textbf{x})$, where $\textbf{x}$ is the vector of inputs, and the covariance (or kernel) function $k(\textbf{x},\textbf{x}')$, which controls the distribution of possible functions that the model can represent.
A GP regression model for some target data $y \sim f(x) + \epsilon$ can be written as $y \sim \mathcal{GP}(m(\textbf{x}), \; k(\textbf{x},\textbf{x}')) + \epsilon$, where $\epsilon$ represents the Gaussian error term common to most regression models.
In this work, we default to the commonly used constant mean GP model with ``squared exponential" or ``radial basis function" (RBF) covariance function: $k(\textbf{x},\textbf{x}') = s^2 \exp(-\frac{1}{2}\|x - x'\| ^2 / \ell^2)$. 
The resulting GP model has several hyperparameters that we tune by gradient-based optimization to maximize the model evidence $p(y) \mid x, \theta)$,
where theta represents the collection of model hyperparameters~\cite{rw}.
The principal hyperparameters for an RBF GP are the amplitude parameter $s$ that controls the overall scale of the model functions, a lengthscale parameter $\ell$ that controls the level of fluctuation relative to the input space, and the observation noise level $\epsilon$.

\paragraph{Acquisition policy: confidence bound + random scalarization}
GP models are commonly used in active learning and optimization settings because they are well-suited to automatic hyperparameter selection and provide good predictive uncertainty estimates.
There are many active learning strategies for selecting experiments based on probabilistic model predictions.
The relative performance of these strategies can strongly depend on the characteristics of the data and the active learning task.
In the optimization examples presented here, we consistently use the confidence bound strategy~\cite{Srinivas2012}, which balances candidate selection between high predicted utility and high predictive uncertainty.
At iteration $t$, an experiment is selected from the design space that minimizes the quantity $\mu(x) - \beta(t) \sqrt{var(x)}$, where $\mu(x)$ represents the mean prediction from a GP model, $\sqrt{var(x)}$  represents the predictive uncertainty, and $\beta(t)$ is an iteration-dependent scaling factor that controls the tradeoff between exploring high-uncertainty regions of parameter space and prioritizing promising candidates for global minima.
We follow~\cite{kandasamy2018} in using the tradeoff schedule $\beta(t) = cd \log(2t + 1)$, where $c = 0.25$ is a fixed hyperparameter of the confidence bound method, and $d$ is the dimensionality of the input space.
We apply the random scalarization~\cite{kandasamy2018} method for weighting competing objectives in a way that encourages fuller exploration of the Pareto frontier.

% Criterion:
% $\mu(x) - \beta_t   \sqrt{var(x)}$
% Confidence bound $c = 0.25$.
% $\beta_t = cd \log(2t + 1)$

\section{Results}\label{sec:results}

\subsection{Multiobjective corrosion property optimization}\label{sec:multiopt}

\begin{figure}[h!btp]
    \centering
    \begin{subfigure}[b]{0.15\textwidth}
         \centering
         \includegraphics[width=\textwidth]{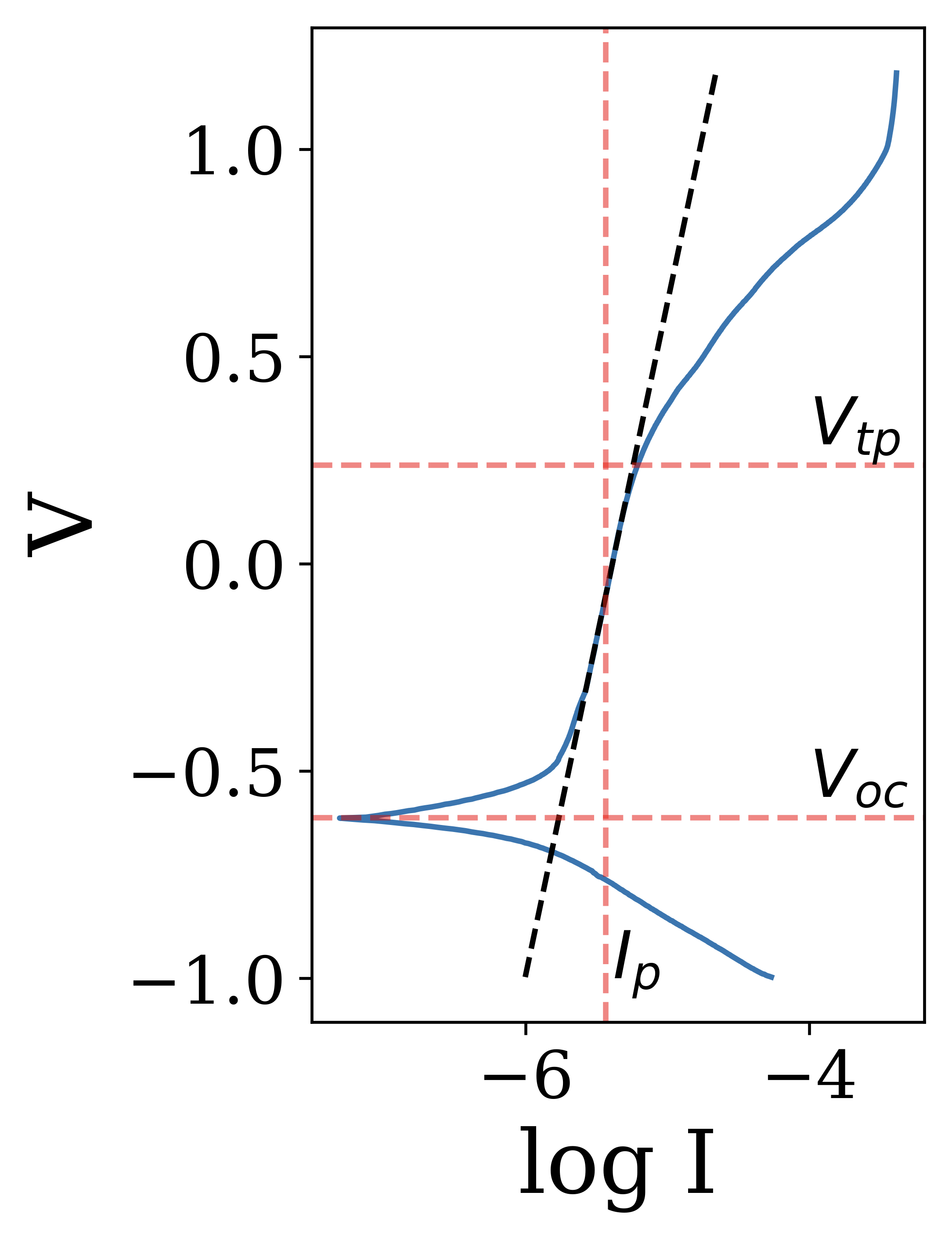}
         \caption{LSV}
         \label{fig:AlNiTi:lpr}
     \end{subfigure}
     \hfill
    \begin{subfigure}[b]{0.22\textwidth}
         \centering
         \includegraphics[width=\textwidth]{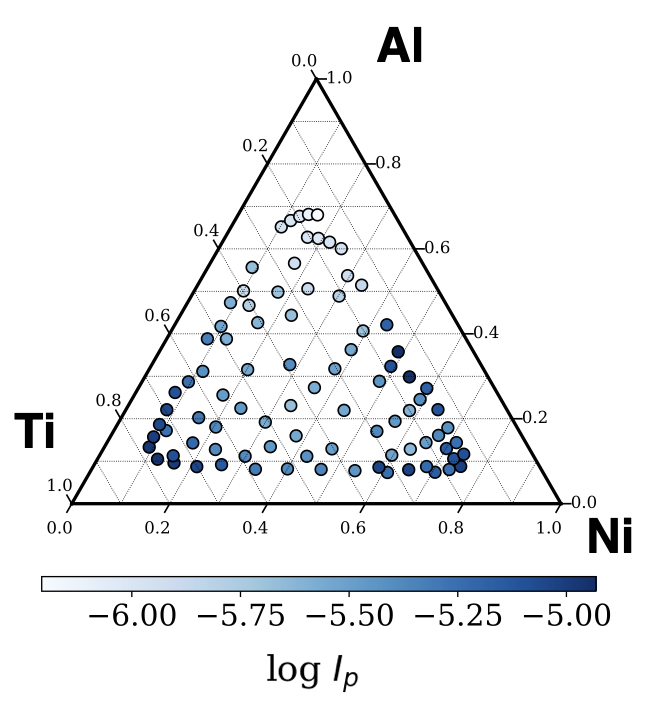}
         \caption{$I_p$}
         \label{fig:AlNiTi:ipass}
     \end{subfigure}
     \hfill
     \begin{subfigure}[b]{0.22\textwidth}
         \centering
         \includegraphics[width=\textwidth]{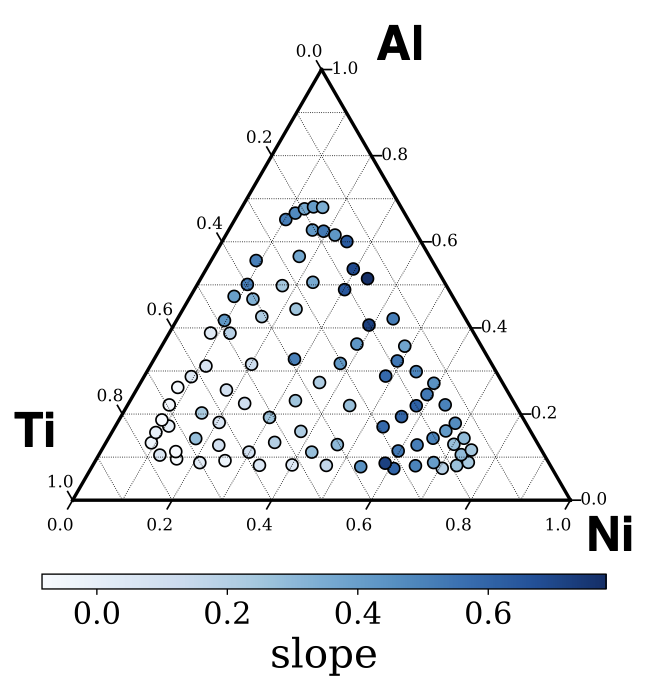}
         \caption{passivation}
         \label{fig:AlNiTi:slope}
     \end{subfigure}
     \hfill
     \begin{subfigure}[b]{0.25\textwidth}
         \centering
         \includegraphics[width=\textwidth]{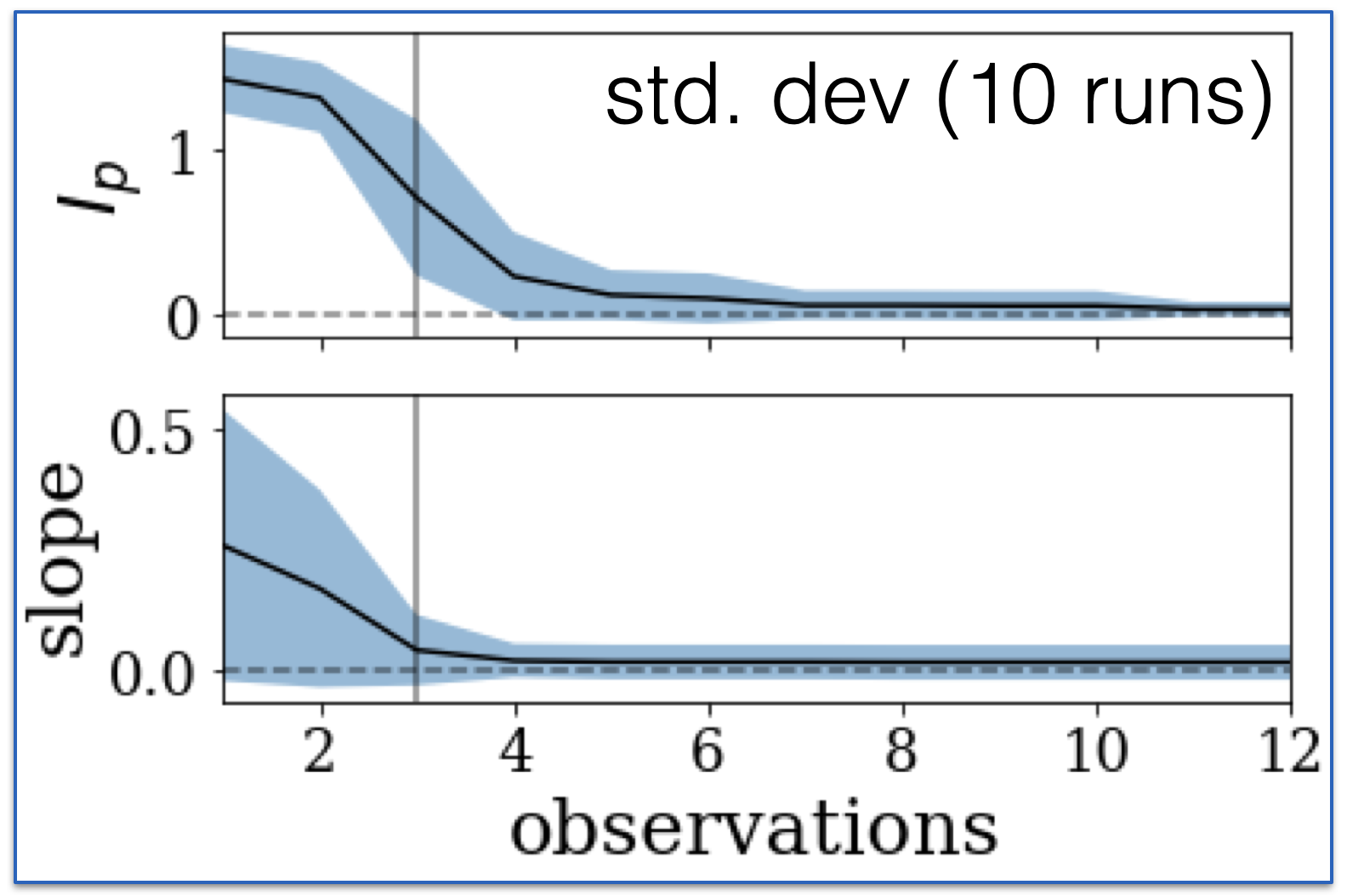}
         \caption{active learning convergence plot.}
         \label{fig:AlNiTi:convergence}
     \end{subfigure}
    \caption{AlNiTi multiobjective case study. (\subref{fig:AlNiTi:lpr}) annotated linear scan voltammetry curve. (\subref{fig:AlNiTi:ipass}) Passivation current vs \ce{AlNiTi} composition. (\subref{fig:AlNiTi:slope}) Passivation plateau slope vs \ce{AlNiTi} composition. (\subref{fig:AlNiTi:convergence}) multiobjective optimization convergence plot.}
    \label{fig:AlNiTi_opt}
\end{figure}

\paragraph{Goal: multiobjective corrosion property optimization / measurement}
Figure~\ref{fig:AlNiTi_opt} shows a multiobjective optimization case study based on a high throughput benchmark dataset of corrosion experiments in a neutral NaCl solution on an AlNiTi composition spread, the details of which can be found in Ref. ~\cite{Joress2020}.
Figure~\ref{fig:AlNiTi:lpr} illustrates a typical  linear scan voltammetry (LSV) result from this series of measurements, showing the measured log current ($I$, amps) as the potential $V$ is swept through an increasing linear schedule.
$V_{oc}$ is the open circuit potential.
We fit a linear model to the passivation plateau at positive potentials relative to $V_oc$; the departure from this linear behavior is the transpassive potential $V_{tp}$.
We determine an effective passivation potential $V_p$ and passivation current $I_p$ by locating the median potential within the linear passivation plateau.
Finally, we characterize the "flatness" of the passivation plateau through the slope of the linear model, $\frac{d \log(I)}{dV}$.

Figures~\ref{fig:AlNiTi:ipass}~and~\ref{fig:AlNiTi:slope} show the measured $I_{p}$ and $\frac{d \log(I)}{dV}$ obtained through SDC corrosion measurements and automated analysis that serves as the benchmark for this case study.  To simulate data collection, we interpolated the data with a GP model and sampled from it based on predetermined acquisition function as described below, adding noise for each simulated measurements. The multiobjective optimization goal is to identify alloys with low passivation current $I_{p}$ and a flat passivation plateau slope (\textit{i.e.,} low $\frac{d \log(I)}{dV}$).

\paragraph{Specific model details}
We model the corrosion response with independent GPs over composition for the two target variables, $\log(I_p)$ and $\frac{d\log(I)}{dV}$.
Each GP uses a constant mean function and an RBF kernel defined over composition variables. 
Measurements are selected with a lower confidence bound strategy~\cite{Srinivas2012} with random scalarization~\cite{kandasamy2018} with objective function weights drawn from $\textrm{Beta}(2,2)$.

%% AlNiTi active learning problem setup and convergence plot

\begin{figure}[h!btp]
    \centering
    \includegraphics[width=0.95\textwidth]{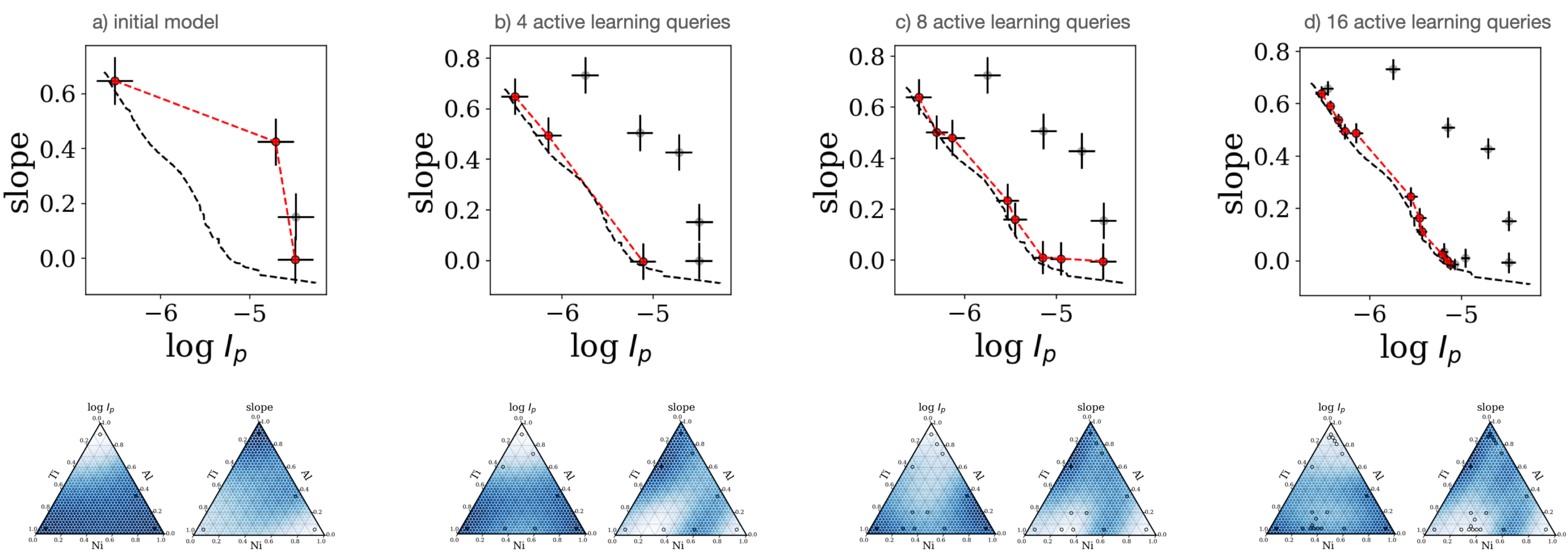}
    \caption{AlNiTi multiobjective active learning trajectory after (a) three initial observations and (b) 4, (c) 8, and (d) 16 active learning queries. The black dashed line indicates the ground truth noise-free Pareto frontier, and the red dashed line indicates the current predicted Pareto frontier. Error bars on observed data indicate the standard deviation of the GP predictive uncertainty.}
    \label{fig:AlNiTiPareto}
\end{figure}

Figure~\ref{fig:AlNiTi:convergence} shows consistent convergence towards both objectives in under 10 active learning queries across a benchmark of ten simulated active learning runs.
We initialize each active learning run with three instances near the corners of the ternary system.
In this system, Al strongly suppressed $I_p$, while Ti is associated with flatter passivation plateau behavior.
As a result, for this particular problem the initial set of measurements tends to be close to optimal for each single objective, as shown by the Pareto plot in the first panel of Figure~\ref{fig:AlNiTiPareto}.
These Pareto plots show the observed values of each objective in a multi-objective optimization problem to succinctly illustrate the set of possible design tradeoffs.

%% AlNiTi pareto

\paragraph{Pareto plot discussion and query allocation}
Figure~\ref{fig:AlNiTiPareto} illustrates the progress of one of these multiobjective optimization trajectories towards exploring the full set of tradeoffs between median passivation current and its range over the passivation plateau.
Though the initial observations are nearly optimal when considering each objective individually, the approximation of the overall Pareto frontier is poor.
However, after four active learning queries (Figure~\ref{fig:AlNiTiPareto}b), the Pareto frontier approximation is within the predictive uncertainty of the GP models.
Subsequent observations concentrate on more densely along the Pareto frontier, and reduce the predictive uncertainty of the GP models as they learn better approximations of the target functions.

\subsection{Semi-mechanistic modeling}\label{sec:semimechanistic}
\paragraph{Problem description and dataset description}
Empirical optimization of material structure and processing is an important research area, but  black box optimization approaches, those without physics embedded in them, are not well-suited for providing the kind of mechanistic insight that drives the development of new materials theory and design heuristics.
This is perhaps the most important challenge facing scalable automated science today: bringing together multiple sources of information to decouple the underlying chemical and structural factors that give rise to the aggregate material properties that we observe.

\begin{figure}[h!btp]
    \centering
    \begin{subfigure}[b]{0.3\textwidth}
         \centering
         \includegraphics[width=\textwidth]{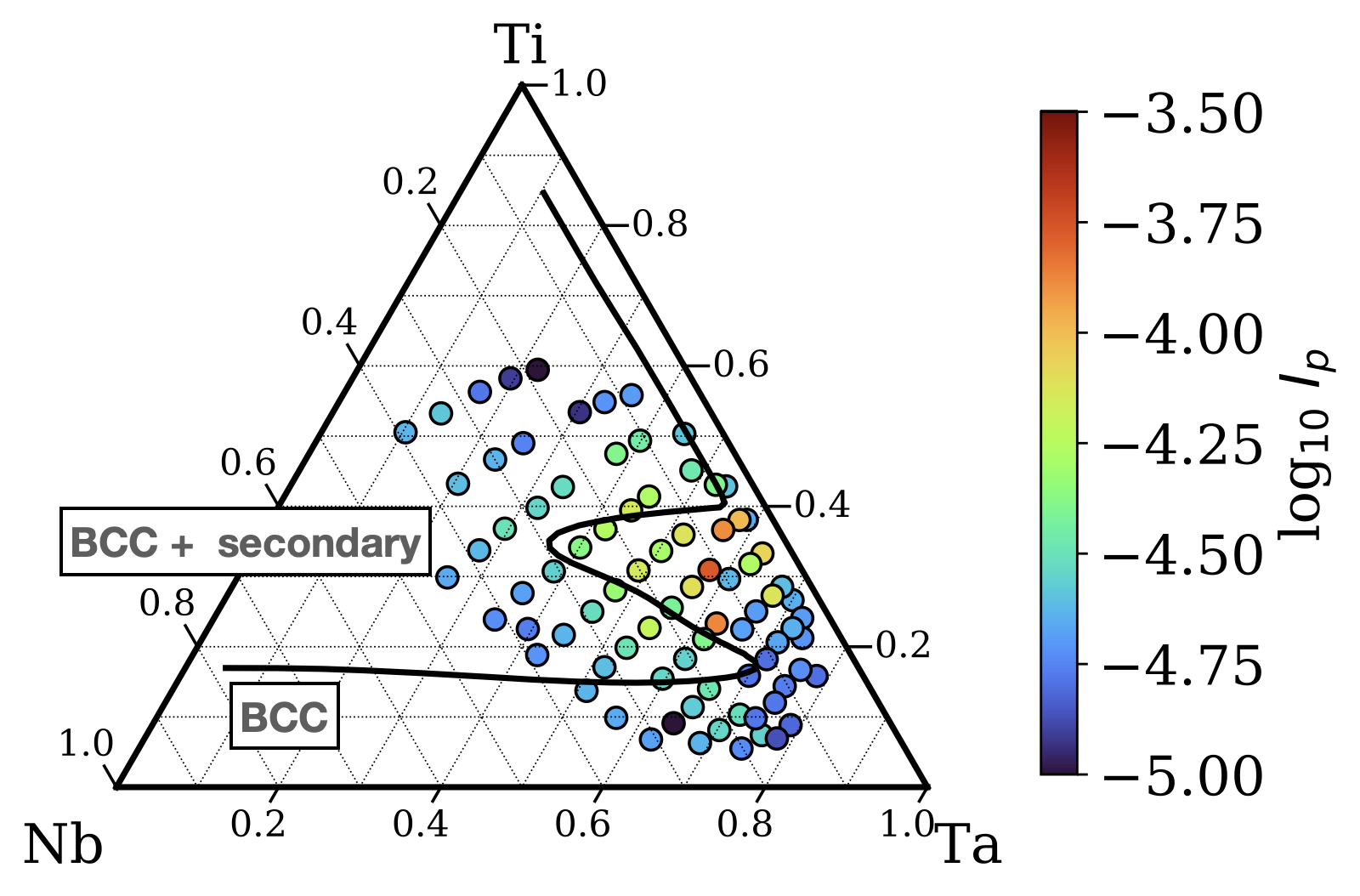}
         \caption{$\log I_{corr}$}
         \label{fig:TiNbTa:current}
     \end{subfigure}
     \hfill
     \begin{subfigure}[b]{0.3\textwidth}
         \centering
         \includegraphics[width=\textwidth]{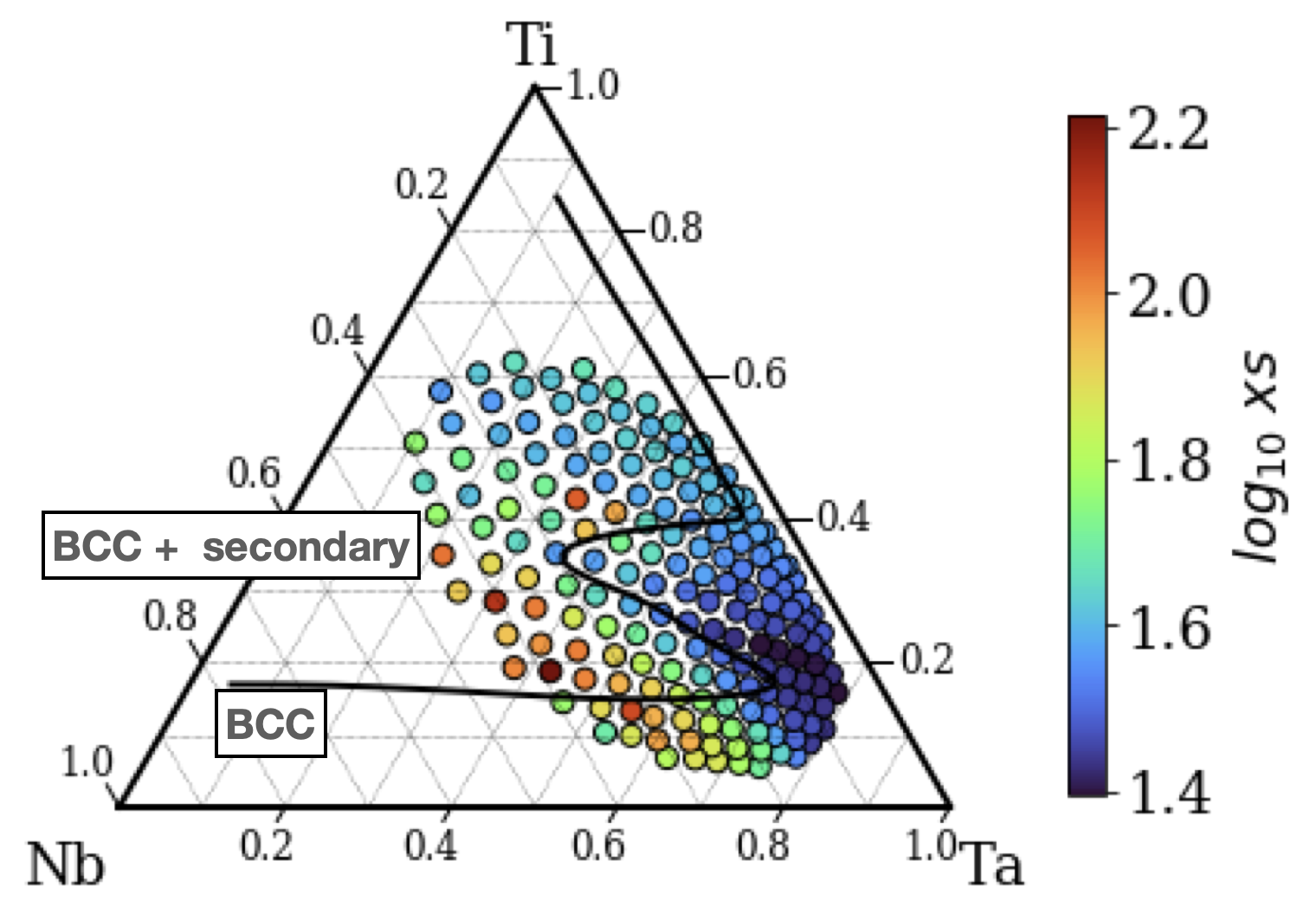}
         \caption{crystallite size}
         \label{fig:TiNbTa:xs}
     \end{subfigure}
     \hfill
     \begin{subfigure}[b]{0.3\textwidth}
         \centering
         \includegraphics[width=\textwidth]{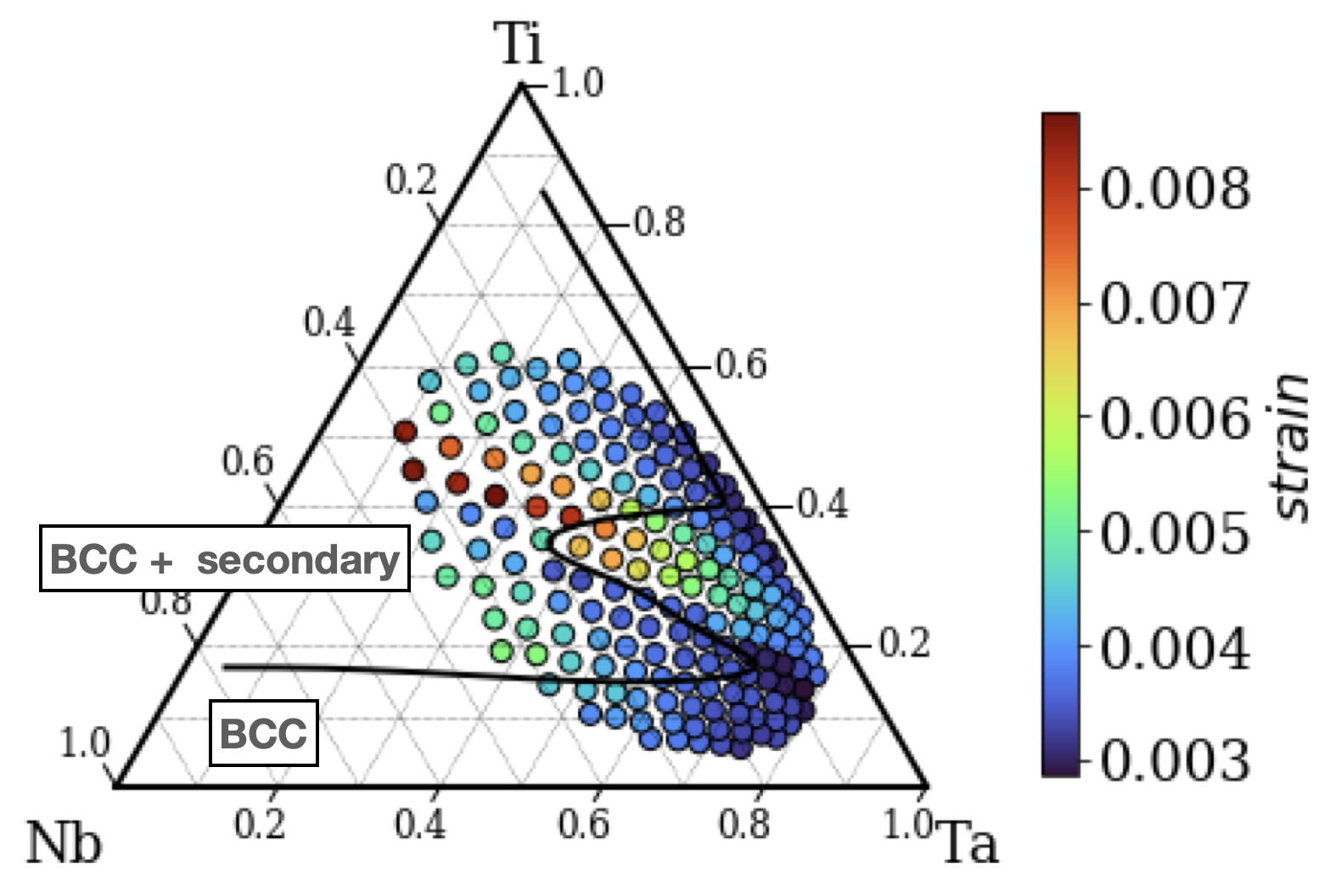}
         \caption{microstrain}
         \label{fig:TiNbTa:mustrain}
     \end{subfigure}

    \caption{TiNbTa (\subref{fig:TiNbTa:current}) corrosion response, (\subref{fig:TiNbTa:xs}) crystallite size, and (\subref{fig:TiNbTa:mustrain}) microstrain.}
    \label{fig:TiNbTa}
\end{figure}

Figure~\ref{fig:TiNbTa} shows (\subref{fig:TiNbTa:current}) the corrosion response (log corrosion current $I_corr$) across a \ce{TiNbTa} composition spread thin film measured by the SDC alongside  (\subref{fig:TiNbTa:xs}) crystallite size and  (\subref{fig:TiNbTa:mustrain}) microstrain, defined here as RMS($\Delta d/d$) where d is the lattice spacing, obtained by Williamson-Hall size-strain analysis~\cite{pecharsky2008} of synchrotron XRD data on the film.
The system nominally forms a body centered cubic (BCC) solid solution, with secondary phase inclusions, determined by subjective inspection of XRD data for unattributed peaks, appearing on the Ta-poor side of the solid black curve.

%% TiNbTa dataset description

Perhaps unexpectedly, the least favorable corrosion rates in Figure~\ref{fig:TiNbTa:current} are found in the single phase, near-equiatomic portion of the ternary system.
The composition dependence of the crystallite size and microstrain suggest that these microstructural features could play a role in modulating corrosion response.
The influence of microstructural features such as grain size~\cite{Ralston2010} and precipitate size and density~\cite{Ralston2010precipitates} is well established.

\paragraph{Model description}
To address these complex mechanistic materials science questions and disentangle potentially competing effects, models with finer granularity are needed.
For example, the partial dependence analysis of the artificial neural network model in~\cite{Zhang2019} provides insight into the contribution of composition and microstructure variables to corrosion behavior.
However, the modeling approach is not amenable to explicit inclusion of theoretically motivated models.

Our approach is to blend theoretical models, such as the Hall Petch grain size contribution of \cite{Ralston2010}, with non-parametric models by specifying additive GPs:

\begin{equation*}
\log(I_{corr}) \sim \mathcal{GP}(\mu, K_{comp}) + k / \sqrt{d}
\end{equation*}

$\mu$ and $K_{comp}$ are a constant mean and standard RBF kernel over composition variables, as in earlier sections of this manuscript.
The grain size $d$ in~\si{nm}  isobtained from size-strain analysis, and $k$ is the linear coefficient in the Hall Petch model, which we optimize along with the other GP hyperparameters.

\begin{figure}[h!btp]
    \centering
    \begin{subfigure}[b]{0.3\textwidth}
         \centering
         \includegraphics[width=\textwidth]{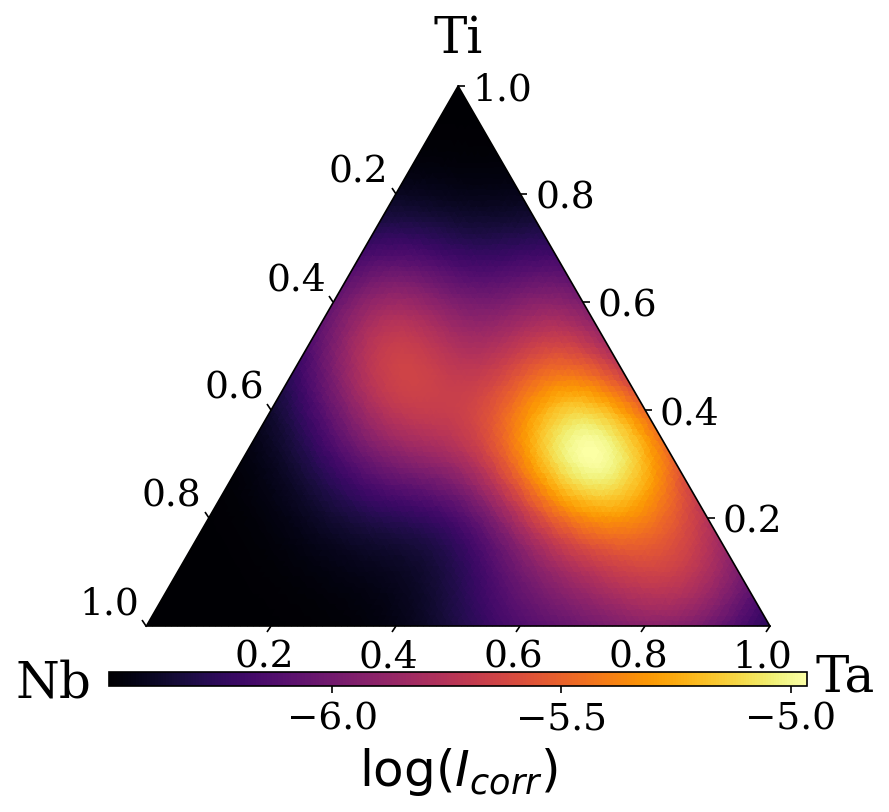}
         \caption{nominal}
         \label{fig:TiNbTa:nominal}
     \end{subfigure}
     \hfill
     \begin{subfigure}[b]{0.3\textwidth}
         \centering
         \includegraphics[width=\textwidth]{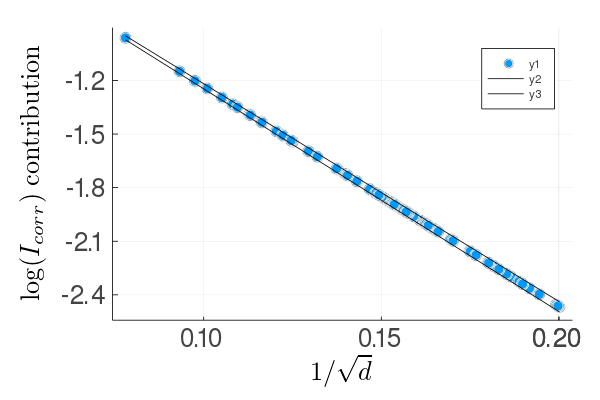}
         \caption{Hall Petch size contribution}
         \label{fig:TiNbTa:xs_effect}
     \end{subfigure}
     \hfill
     \begin{subfigure}[b]{0.3\textwidth}
         \centering
         \includegraphics[width=\textwidth]{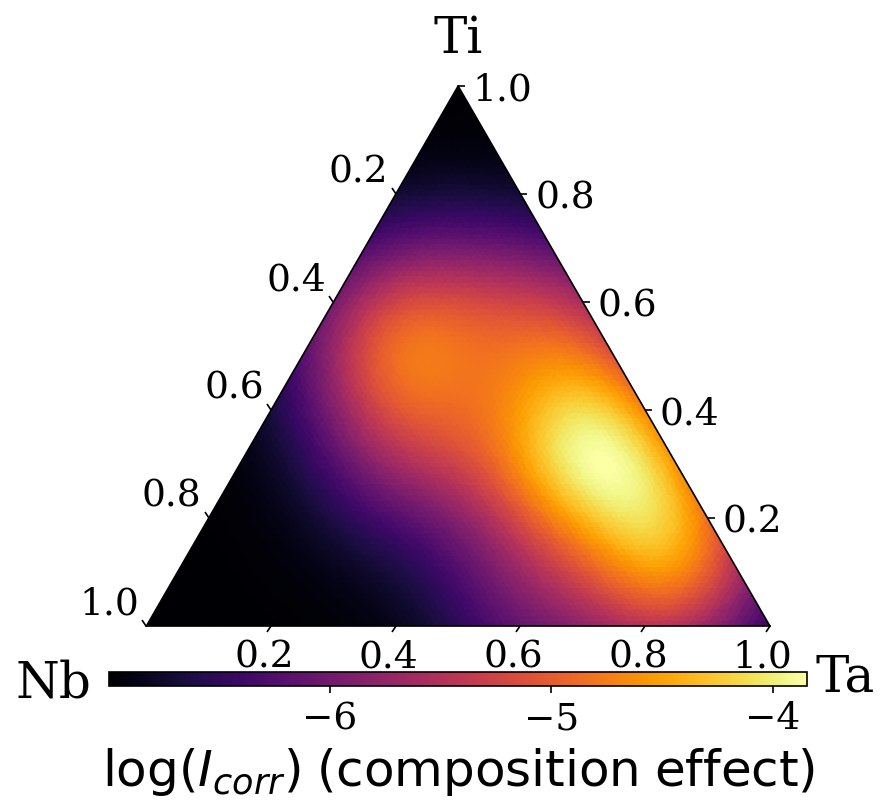}
         \caption{Composition contribution}
         \label{fig:TiNbTa:composition_effect}
     \end{subfigure}

    \caption{(\subref{fig:TiNbTa:nominal}) Model $\log(I_{corr})$ response with decoupled (\subref{fig:TiNbTa:xs_effect}) grain size and (\subref{fig:TiNbTa:composition_effect}) composition contributions.}
    \label{fig:TiNbTaresults}
\end{figure}

\paragraph{semimechanistic model results}
Figure~\ref{fig:TiNbTa:nominal} shows the nominal $\log(I_{corr})$, along with the decoupled (\subref{fig:TiNbTa:xs_effect}) Hall Petch contribution and (\subref{fig:TiNbTa:composition_effect}) composition contribution with grain size set to 40~\si{nm}.
The composition contribution spans a larger dynamic range of  $\log(I_{corr})$ values, and the location of the maximum value has shifted and spans a larger composition range in comparison to the nominal model.

\section{Discussion}\label{sec:discussion}

\paragraph{automating scientific goals}
Composing a complex, semi-mechanistic model as in Section~\ref{sec:semimechanistic} allows us to explore counterfactual predictions like ``what if we could independently increase/decrease the grain size", as in Figure~\ref{fig:TiNbTa:composition_effect}.
Similarly, competing mechanistic model components could automatically be evaluated, for example with an alternative Hall Petch type model.
With the present composition spread dataset, this model is merely descriptive, and it is difficult to validate these kinds of mechanistic hypotheses without the ability to independently vary composition and microstructure variables.  Creating datasets that can fully span the range of composition and parameter space is in general intractable with conventional experimental synthesis methods and even high-throughput methods.
Meeting this challenge, therefore, represents a huge growth opportunity for autonomous materials science platforms.
What is needed are agile automated platforms that 1) use semi-mechanistic models that provide quantitative insight into underlying material attributes that drive behavior, 2) can automatically and with low latency synthesize material with the desired attributes, 3) integrate enough online characterization streams to inform all relevant model components, and 4) use novel planning algorithms that use scientific criteria to plan experiments to optimally evaluate mechanistic hypotheses.

\paragraph{semi-mechanistic models and complexity}
The nascent field of scientific ML~\cite{baker2019workshop} will play a strong role in expanding the capability of autonomous materials platforms to broader scientific inquiry.
One of the specific challenges to address in materials applications is the sometimes overwhelming complexity of interacting chemical and structural processes that mediate materials behavior.
Consider the simple additive GP model in the TiNbTa vignette (Section~\ref{sec:semimechanistic}).
The model explicitly attempts to account for grain size effects, but many other important structural characteristics are absorbed by the non-parametric GP over composition.
The crystal structure of the primary phase and presence of secondary phases (and their structure, volume fraction, size distribution, \textit{etc.}) are expected to play a large role on corrosion resistance and other important material properties.
Similarly, additional microstructural features like crystallographic texture and defect content and character can heavily influence material properties.
Some of the aspects of materials structure have well-developed theoretical frameworks, while others may be treated in a largely empirical fashion.
A major goal of materials-oriented scientific ML research should therefore be to integrate many different materials modeling types and approaches into a flexible ML framework that can bridge the gaps between mechanistic insight and empiricism.

\paragraph{Challenges in learning to make things}
To actually parameterize such models in an online fashion, continued creativity and innovation in automated online synthesis capability is needed to actually make suitable samples.
This may constitute an active learning task of its own.
For example, \textit{a priori} specification of alloy electrodeposition conditions to obtain a targeted composition is a challenging task.
There are competing thermodynamic and kinetic factors at play, and while there is a rich landscape of theoretical models, it is difficult to predict which mechanism will dominate, especially in a previously unexplored system.
This is further complicated by a need to learn how to control microstructural factors like grain size, texture, and phase distribution independently from alloy composition, to the extent possible.

\begin{figure}[h!btp]
    \centering
    \begin{subfigure}[b]{0.3\textwidth}
         \centering
         \includegraphics[width=\textwidth]{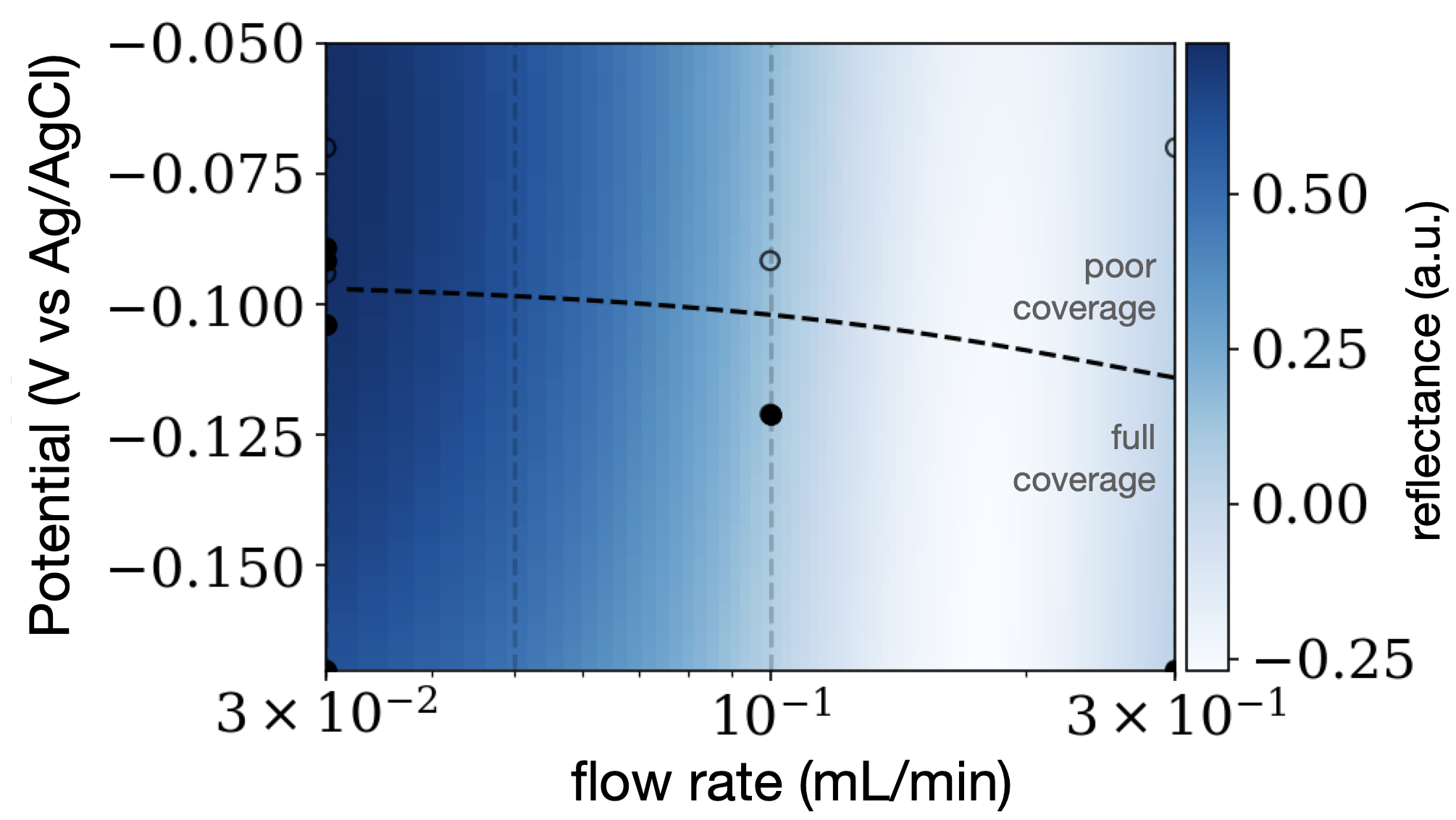}
         \caption{reflectance}
         \label{fig:deposition:Cureflectance}
     \end{subfigure}
     \hfill
     \begin{subfigure}[b]{0.3\textwidth}
         \centering
         \includegraphics[width=\textwidth]{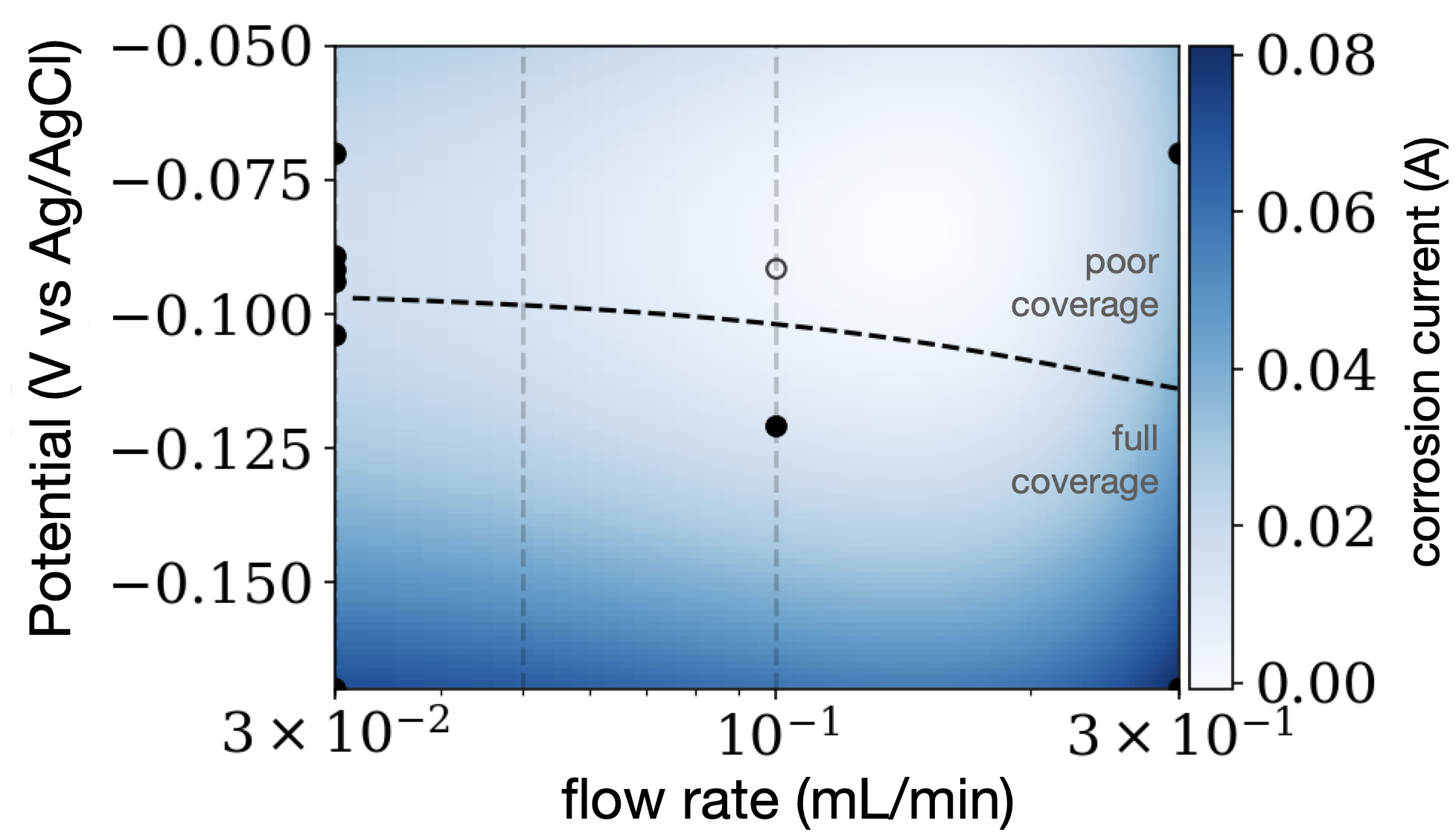}
         \caption{corrosion current}
         \label{fig:deposition:Cucurrent}
     \end{subfigure}
     \hfill
     \begin{subfigure}[b]{0.3\textwidth}
         \centering
         \includegraphics[width=\textwidth]{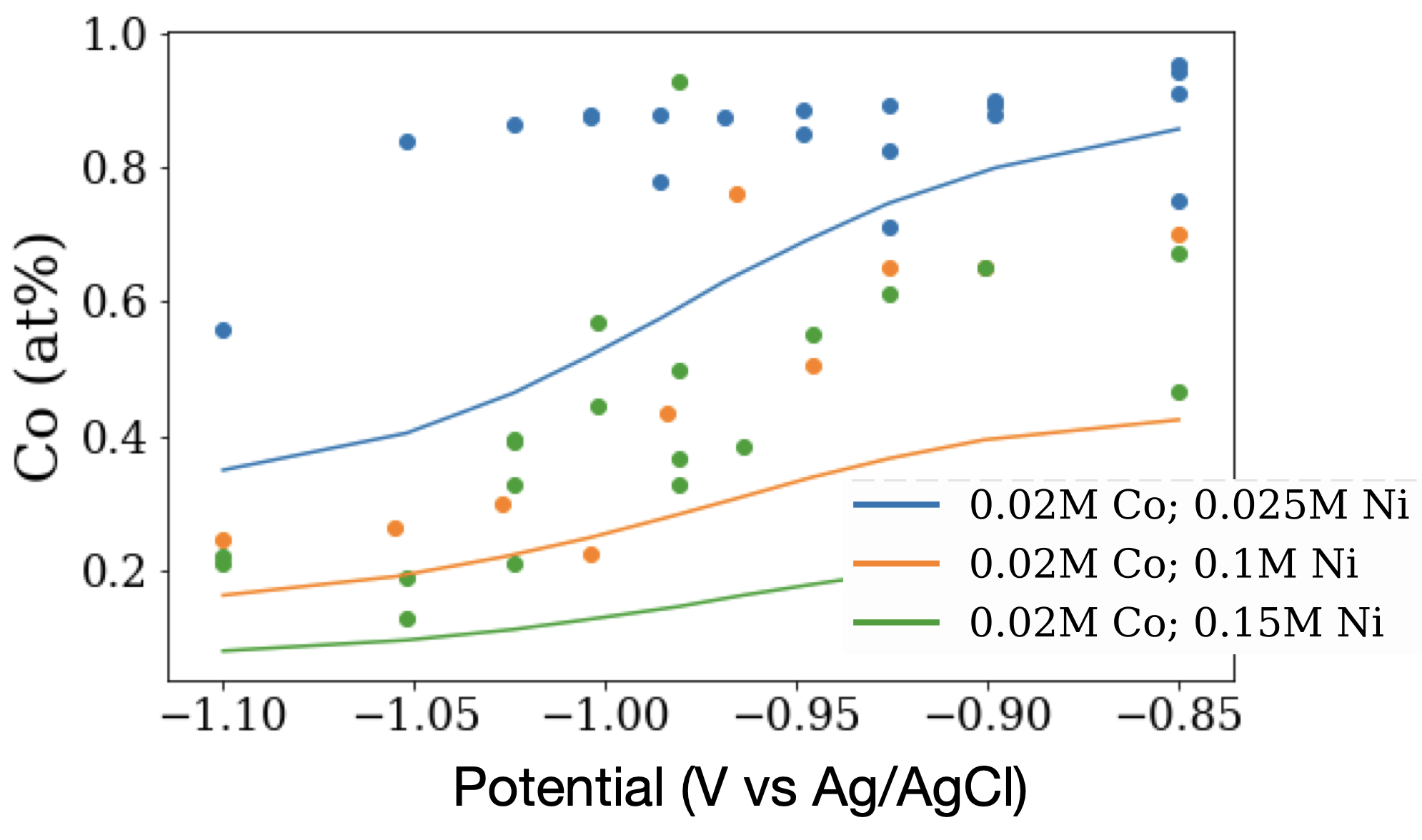}
         \caption{NiCo series}
         \label{fig:deposition:NiCo}
     \end{subfigure}

    \caption{Preliminary alloy electrodeposition studies. (\subref{fig:deposition:Cureflectance}) and (\subref{fig:deposition:Cucurrent}) show the modeled laser reflectance and corrosion current of deposited Cu films after ten active learning iterations; the dashed line shows a GP classification model for deposit coverage. (\subref{fig:deposition:NiCo}) shows measured vs expected electrodeposited NiCo alloy composition as a function of solution composition and applied potential during deposition.}
    \label{fig:deposition}
\end{figure}

Figure~\ref{fig:deposition} illustrates these issues using  electrodeposition data from (\subref{fig:deposition:Cureflectance}) and (\subref{fig:deposition:Cucurrent}) Cu metal and  (\subref{fig:deposition:NiCo}) preliminary alloy electrodeposition studies. 
The goal of the copper study is to learn the feasible processing envelope, depositing copper metal for 5 minutes at constant potential from 0.25~\si{\Molar} \ce{CuSO4} onto a gold thin film.
The GP confidence bound method is used to optimize (\subref{fig:deposition:Cureflectance}) the laser reflectance of the deposit (as in~\ref{fig:BNL:setup}b) as a proxy for roughness, and to minimize the average dissolution current over a two minute constant potential hold at 100~\si{mV} vs \ce{Ag}/\ce{AgCl} reference in 1~\si{\Molar} \ce{H2SO4}.
The dashed line shows the predictions of a GP classification model for deposition coverage (sufficient/poor); the classification uncertainty is included in the active learning acquisition policy.

The NiCo alloy deposition case study in Figure~\ref{fig:deposition:NiCo} illustrates some of the challenges in electrodepositing alloys with targeted composition.
Alloy were deposited at varying constant potential setpoints in solutions of 0.02~\si{\Molar} \ce{CoSO4}; 0.025~\si{\Molar} \ce{NiSO4} (blue),  0.02~\si{\Molar} \ce{CoSO4}; 0.1~\si{\Molar} \ce{NiSO4} (orange), and 0.02~\si{\Molar} \ce{CoSO4}; 0.15~\si{\Molar} \ce{NiSO4} (green).
Alloy compositions were determined offline though energy dispersive spectroscopy (EDS) line scans across each deposit.
The solid curves show expected alloy composition based on a simple linear deposition current combination model based on single-component potential-current calibration curves.
The EDS data span nearly the full binary composition range, but show substantial deviation from the naive expectations, including near insensitivity to deposition potential at the lowest solution concentrations.
With online composition feedback, an active learning system may be able to quickly learn the nonlinear interactions between solution composition and deposition driving force to obtain desired alloy compositions.

\paragraph{measuring and quantifying the relevant quantities}
In addition to developing innovative on demand synthesis technologies, for maximum impact these platforms need to be tightly integrated with a diverse array of materials characterization technologies.
For example, in the NiCo case study, we not only want to learn to electrodeposit alloys with high quality surfaces, but we may wish to try to independently control grain size, or target specific structural phases (either stable or metastable).
Conventionally, the materials synthesis and characterization tools needed to produce all this data are decoupled from each other, and temporal latency between steps is high.
Modularization and miniaturization are therefore a high priority for expanding the scope and impact of automated materials science.
Wherever modular integration of critical characterization is not yet feasible (\textit{e.g.,} transmission electron microscopy), adopting and improving batch active ML algorithms will be important for integrating this information into research feedback loops\cite{ren2018accelerated}.
Additionally, quantitative high throughput measurement of many important materials properties presents yet another category of important scientific ML sub-problems.
For example, automated high throughput phase identification and phase fraction analysis is still a major challenge, particularly in systems where minor and trace levels of secondary phases can potentially play a large role on effective material properties.

\paragraph{agile, science-based experiment planning tools}
Finally, with interpretable scientific ML algorithms and the capability to make and characterize samples on demand, advances in scientific planning algorithms are ripe for high impact.
An important first step might include algorithms for planning experiments to identify which latent structural factor is driving material performance.
This could be coupled with model visualization and interrogation tools that allow scientists to formulate mechanistic hypotheses in terms of available modeling components, and to reformulate mechanistic and semi-mechanistic models throughout the data acquisition process.
A more ambitious task is to jointly learn how to synthesize materials with the necessary chemical and structural attributes to adaptively perform a series of experiments driven by a scientific model selection criterion.
Similarly, automated planning systems could use competing semi-mechanistic models to map out composition and processing ranges where a particular physical mechanism is operative, even when some model components need to be treated empirically.

\section{Conclusion}\label{sec:conclusion}

General autonomous science systems face several challenges: learning to reliably synthesize materials, mapping material specification and processing to structure and properties, incorporating offline data streams, and incorporating prior theoretical and data-driven knowledge.
As the materials community surmounts these challenges, closed-loop automated materials synthesis and characterization platforms offer much more than a means of engineering materials properties and performance through black-box optimization algorithms: they offer the potential to develop and deploy new algorithms for generating and testing scientific hypotheses.

\backmatter

\bmhead{Supplementary information}

\bmhead{Acknowledgments}
BD and HJ acknowledge partial support from the National Research Council Research Associate Program.

\section*{Declarations}
On behalf of all authors, the corresponding author states that there is no conflict of interest.

Data availability: The raw data required to reproduce the analyses in this manuscript will be made available in a public data repository.

Code availability: The automation software for the scanning droplet cell platform will be made available at \url{https://github.com/USNISTGOV/auto-sdc}. Code required for reproducing the analyses in the manuscript will be made available in a separate repository.

\bibliography{sn-bibliography}% common bib file


\end{document}